\documentclass[conference,10pt]{IEEEtran}
\usepackage{amsmath}
\usepackage{amssymb}
\usepackage{epsfig}
\usepackage{cite}
\usepackage{multirow}
\usepackage{dsfont}
\usepackage{graphicx}
\usepackage{array}
\usepackage[hyphens]{url}
\usepackage{mathtools}
\usepackage{algorithmic}
\usepackage{algorithm}
\usepackage{mathtools}
\usepackage{balance}
\usepackage{seqsplit}

\usepackage{multicol}
\usepackage{listings}

\usepackage{color}

\newcommand{\red}[1]{\textcolor{red}{\bf #1}}

\newcommand{\warn}[1]{}
\newcommand{\nop}[1]{}

\begin{document}

\title{IoT Security: An End-to-End View and Case Study}

\author{\IEEEauthorblockN{Zhen Ling\IEEEauthorrefmark{1}, Kaizheng Liu\IEEEauthorrefmark{1}, Yiling Xu\IEEEauthorrefmark{1}, Chao Gao\IEEEauthorrefmark{5}, Yier Jin\IEEEauthorrefmark{2},  Cliff Zou\IEEEauthorrefmark{3}, Xinwen Fu\IEEEauthorrefmark{3}, and Wei Zhao\IEEEauthorrefmark{4}
\IEEEauthorblockA{\IEEEauthorrefmark{1}Southeast University,
Email: \{zhenling, kzliu, ylxu\}@seu.edu.cn}
\IEEEauthorblockA{\IEEEauthorrefmark{5}University of Massachusetts Lowell,
Email: cgao@cs.uml.edu}
\IEEEauthorblockA{\IEEEauthorrefmark{2}University of Florida,
Email: yier.jin@ece.ufl.edu}
\IEEEauthorblockA{\IEEEauthorrefmark{3}University of Central Florida,
Email: \{Changchun.Zou, xinwenfu\}@ucf.edu}
\IEEEauthorblockA{\IEEEauthorrefmark{4}University of Macau, China,
Email: weizhao@umac.mo}
}}


\maketitle
\pagestyle{plain}

\pagenumbering{arabic}

\begin{abstract}
In this paper, we present an end-to-end view of IoT security and privacy and a case study. Our contribution is three-fold. First, we present our end-to-end view of an IoT system and this view can guide risk assessment and design of an IoT system. We identify 10 basic IoT functionalities that are related to security and privacy. Based on this view, we systematically present security and privacy requirements in terms of IoT system, software, networking and big data analytics in the cloud. Second, using the end-to-end view of IoT security and privacy, we present a vulnerability analysis of the Edimax IP camera system. We are the first to exploit this system and have identified various attacks that can fully control all the cameras from the manufacturer. Our real-world experiments demonstrate the effectiveness of the discovered attacks and raise the alarms again for the IoT manufacturers. Third, such vulnerabilities found in the exploit of Edimax cameras and our previous exploit of Edimax smartplugs can lead to another wave of Mirai attacks, which can be either botnets or worm attacks. To systematically understand the damage of the Mirai malware, we model propagation of the\nop{ original} Mirai\nop{and an enhanced one} and use the simulations to validate the modeling. The work in this paper raises the alarm again for the IoT device manufacturers to better secure their products in order to prevent malware attacks like Mirai.
\end{abstract}




\section{Introduction}
\label{sec::Introduction}

IoT can be defined as interconnecting various uniquely addressable objects through communication protocols. 
IoT is booming and changing our life. We can interconnect anything including virtual objects together and access those things remotely \cite{AIM::IoT::2010, GBMP::IoT::2013}. IoT has broad applications, including healthcare, life sciences, municipal infrastructure, smart home, retail, manufacturing, agriculture, education and automation. Forbes reported that by 2020 annual revenue of IoT vendors could exceed \$470B, Industrial Internet of Things (IIoT) will exceed 60 trillion in the next 15 years and IoT market size was about 900M in 2015 \cite{Columbus::IoTForecasts::Forbes2016}. According to Gartner's hype cycle of emerging technologies in 2016 \cite{Gartner::Hype::2016}, the expectation for IoT is very high and standardization of IoT platforms will need 5-10 years. The IEEE P2413 Working Group has been trying to standardize the IoT framework since 2014 while there is no consensus yet \cite{IEEE::P2413::2016} .

IoT has attracted hackers. There are two kinds of threats: threats against IoT and threats from IoT. 1. {\it Threats against IoT}: On Oct. 21, 2016, a huge DDoS attack was deployed against Dyn DNS servers and shut down many web services including Twitter \cite{Hil::Dyn::2016}. Hackers exploited default passwords and user names of webcams and other IoT devices, and installed the Mirai botnet \cite{Antonakakis+::MiraiBotnet::Security2017} on compromised IoT devices. The huge botnet was then used to deploy the DDoS attack against Dyn DNS servers. Various other IoT devices have been hacked. IP cameras can be hacked through buffer overflow attacks \cite{Chi::CCTV::2016}. Philips Hue lightbulbs were hacked through its ZigBee link protocol \cite{Rab::Israeli::2016}. SQL injection attacks were effective against Belkin IoT devices \cite{Kov::Belkin::2016}. 2. {\it Threats from IoT}: Researchers also find cross-site scripting (XSS) attacks that exploited the Belkin WeMo app and access data and resources that the app can access \cite{Kov::Belkin::2016}. A IoT enabled UAV can fly far away and compromise the security and privacy of people on ground.

In this paper, we focus on safeguarding IoT and present an end-to-end view of IoT security and privacy. Our contribution is three-fold. First, we present our end-to-end view of an IoT system. We identify 10 basic IoT functionalities related to security and privacy while an IoT system may not have all these 10 functionalities. Based on this view, we systematically analyze the security and privacy requirements in terms of five dimensions: hardware, operating system/firmware, software, networking and big data analytics in the cloud. Second, using the end-to-end view of IoT security and privacy, we present an attack against the Edimax IP camera system. We are the first to exploit this system and have identified various attacks that can fully control all Edimax cameras of the model of interest. The exploit of the camera system demonstrates the usefulness of our view of IoT security and privacy. Third, the attacks against Edimax cameras and our previous attacks against Edimax smartplugs \cite{LLX+::SmartPlug::IoTJ17} allow attackers to install the Mirai malware onto those systems. To evaluate the damage of the Mirai malware systematically, we model the prorogation of the Mirai\nop{ and an enhanced one}. The Mirai propagation is different from the prorogation of traditional worms since it adopts a password dictionary attack against target IoT devices. The dictionary attack affects the prorogation speed. We use NS3 simulations to validate the modeling. Our model also matches the real-world data of Mirai.

A conference version of this paper is included in the proceedings of IEEE GLOBECOM 2017 \cite{Ling+::SecurityPrivacy::Globecom2017}. Compared with the conference version, this journal version provides more details of our end-to-end view of IoT security and privacy in Section \ref{subsec::IoTSP}, particularly on microcontrollers (MCUs). We introduce more attacks against the IP camera in Section \ref{subsec::OtherAttacks}. Mirai modeling is provided in Section \ref{sec::propagation} and simulation results are presented in Section \ref{subsec::Mirai}.

The rest of the paper is organized as follows. We introduce our end-to-end view of IoT security and privacy in Section \ref{sec::View}. In Section \ref{sec::Protocol}, we introduce the communication protocol of the Edimax camera system. Then we present our exploit of the Edimax camera system in Section \ref{sec::vulnerabilities}. The\nop{ original and enhanced} Mirai propagation models are presented in Section \ref{sec::propagation}. We evaluate the exploit and the propagation of Mirai and compare our model with real world data in Section \ref{sec::Evaluation}. Section \ref{sec::Conclusion} concludes the paper.

\section{Framework of IoT Security and Privacy}
\label{sec::View}
In this section, we first present our end-to-end view of an IoT system and then present security and privacy requirements for an IoT system.

\subsection{End-to-End View of IoT}
\label{subsec::e2eiot}
\begin{figure}[htp]
\centering
\includegraphics[height=1.8in]{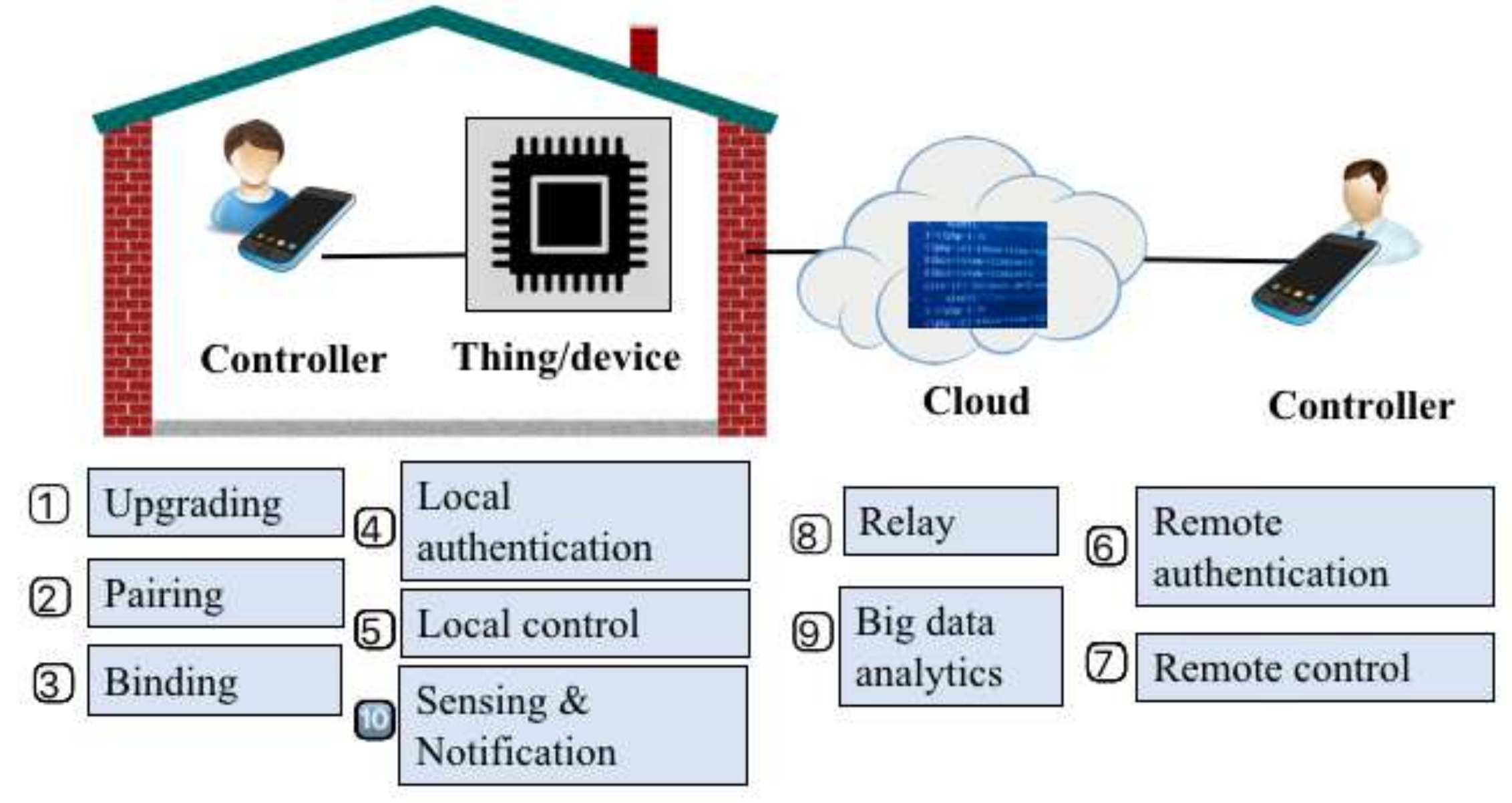}
\vspace{-0.2cm}
\caption{End-to-end View of IoT}
\label{fig::IoT}
\vspace{-0.4cm}
\end{figure}

We will focus on a standalone IoT system as shown in Figure \ref{fig::IoT}. Such a system normally has three basic components: {\it thing}, {\it controller} and {\it cloud}. The thing is connected to the Internet. For a smart home system, the {\it thing} is normally behind a wireless router, which adopts NAT to set up a local network of home systems. The {\it controller} can be a program on a PC or app on a smart device such as a smartphone or tablet. Without loss of generality, we often use a smartphone as an example controller in this paper. Within the local network, the controller can communicate with the thing through the router. However, if the controller is outside, it will not be able to contact the thing directly since the thing is behind NAT (unless port forwarding is enabled on the home router for the thing). Therefore, most IoT systems use a cloud as an intermediate relay between the thing and controller. The thing builds a permanent connection to the cloud. The controller controls or requests information from the thing through the cloud.

We have identified 10 basic functionalities in a IoT system. Please note an IoT system may not have all these 10 functionalities because of its design goal and application.
\begin{enumerate}
\item {\bf Upgrading}: The firmware of the thing can be upgraded to provide more and better services or a security patch can be applied. The firmware can be a full-fledged embedded Linux system. If the thing is a microcontroller (MCU), the firmware can be a piece of dedicated code for simple control or sensing. For example, a MCU can be used to turn on and off an air conditioner.
\item {\bf Pairing}: Bootstrapping a thing generally involves two steps, pairing and then binding. A controller like a smartphone should be able to communicate with a thing at the bootstrapping time. Such a communication channel can be WIFI, Bluetooth, ZigBee, barcode/scanner, and near field communication. This {\it connecting process} is denoted as pairing. For example, when many smart things on market are powered on, they behave as a wireless router and allow the controller to connect to the things in order to configure the things. Apparently, if a thing is deployed in public, we have to limit who can access the thing and configure it.
\item {\bf Binding}: This is the process of configuring the thing through the controller once pairing is done. The controller may bind the thing to the Internet, that is, connect the thing to the Internet. For example, the controller can require a user to input the WiFi SSID (Service Set Identifier) and password of a wireless router and send the information to the thing, which can then connect to the Internet. Another important binding activity is to bind the thing and its users. For example, the controller can learn the identity of the thing (e.g., the MAC address of the wireless interface on the thing) via the communication channel used in the pairing process. Therefore, the user and the thing can be bound together via an appropriate protocol.  
\item {\bf Local authentication}: Within a local network, the controller may connect to a port open on a thing, which should authenticate the user and then allow further actions from the user.
\item {\bf Local control}: Once a user is authenticated, the controller can send commands to control the thing.
\item {\bf Remote authentication}. If the controller is on the Internet and not in the local network, it may not be able to directly contact the thing, which may be behind NAT, and has to go through the cloud for authentication.
\item {\bf Remote control}: If the controller is on the Internet and not in the local network, it may have to control the thing through a cloud.
\item {\bf Relay} by cloud: For remote authentication and control, the cloud is to relay the authentication and control messages between the thing and controller. The cloud may have an authentication server to authenticate both the thing and controller and connect them together.
\item {\bf Big data analytics} by cloud: The cloud may collect the data from things and users, and perform big data analytics. A cloud may connect to other clouds that serve other things, share data and request further analytics capabilities.
\item {\bf Sensing and notification}: Many things are smart. For example, a thing may sense the room temperature and notify the user if the temperature is too low or high. A thing can also notify the user about abnormal behaviors such as too many login attempts on the thing.
\end{enumerate}

\subsection{Security and Privacy in IoT}
\label{subsec::IoTSP}
\begin{figure}[htp]
\begin{center}
\includegraphics[height=1.8in]{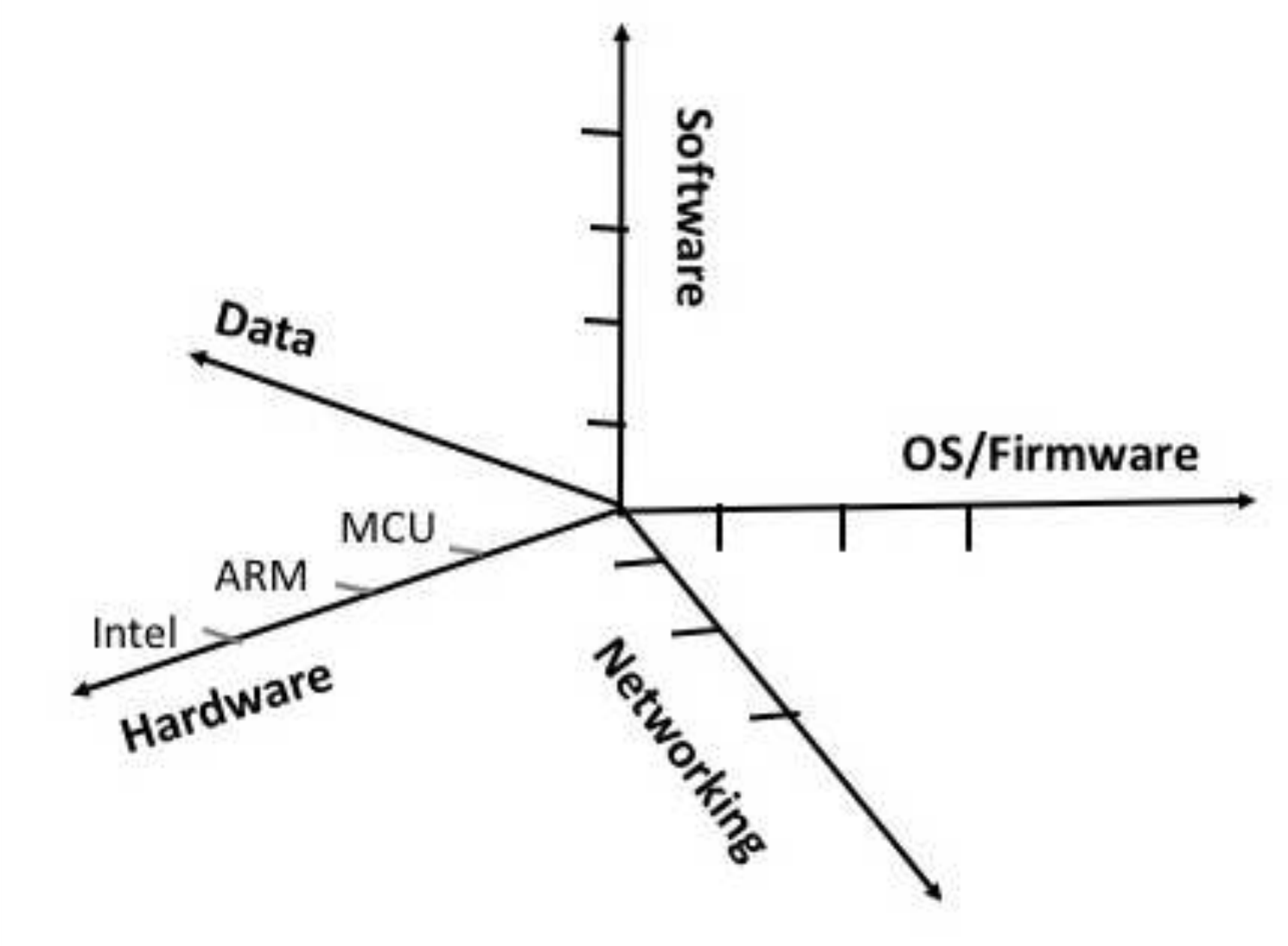}
\vspace{-0.2cm}
\caption{Five Aspects of IoT Security and Privacy}
\label{fig::IoT5}
\end{center}
\vspace{-0.3cm}
\end{figure}

To secure an IoT system, we have to consider five dimensions: hardware, operating system/firmware, software, networking and data generated and maintained within the system, as shown in Figure \ref{fig::IoT5}. As illustrated in Figure \ref{fig::IoT}, an IoT system has quite a few components, all of which should be inspected from these five aspects. The 10 functionalities of IoT identified in Section \ref{subsec::e2eiot} span across these five dimensions. We have to secure any interface that may interact with users (including attackers) in an IoT system.

{\bf Hardware security}:
Hardware security is critical when attackers can physically access the IoT devices.
For example, many IoT devices do not disable their debugging ports after the testing and validation stage, which give attackers full access to the internal firmware. In fact, almost all IoT devices have hardware vulnerabilities which may be exploited by attackers including  the UART/JTAG debugging ports, multiple boot options, and unencrypted flash memory \cite{Arias+::Wearable::TMSCS2015,Wurm+::Obfuscation::ASP-DAC2016,Jin+::NestThermostat::BlackHat2014}.
Through the hardware backdoors, attackers can easily bypass software level integrity checking by either disabling the checking functionality or booting the system through an injected firmware image. An IoT security vulnerability database is recently constructed, which presents a large spectrum of different types of vulnerabilities including hardware security related vulnerabilities \cite{ucf::IoTVulnerabilityDatabase::2017}. Accordingly, countermeasures are recently proposed to prevent physical attacks such as runtime attestation to prevent TOCTOU attacks \cite{Zeitouni+::ATRIUM::ICCAD2017}. TPM \cite{Tom::TPM::2008}, TrustZone \cite{ARM::TrustZone::2009} and Intel SGX \cite{Intel::SGX::2017} can provide hardware-level security.

Microcontrollers (MCUs) have been broadly used in industry including automobiles and home automation. For example, the Melzi board uses ATmega chips and is a very popular 3D printer control board. We use one popular 8-bit microcontroller ATmega1284P to demonstrate the security issues in MCUs. ATmega1284P is a low-power CMOS 8-bit microcontroller based on the AVR enhanced RISC architecture. It can run instructions in a single clock cycle and achieves throughputs around 1MIPS per MHz.
ATmega1284P has 128KB in-system programming (ISP) flash memory, 16KB SRAM, 4KB EEPROM, 32 General Purpose Working Registers, two USARTs, SPI serial port, and a JTAG (IEEE 1149.1 compliant) test interface for on-chip debugging and programming.
{\em Fuse} and {\em lock} bits can be used to lock writing and reading to/from the flash memory from either the application area or the bootloader section through JTAG/SPI or other ports. These bits can also lock writing and reading to/from the EEPROM memory through JTAG/SPI or other ports. Programming the fuse and lock bits will protect the content written to the on-chip flash and EEPROM memory.
Setting fuse bits can disable only the JTAG/OCDEN/SPI ports and setting lock-bits can prevent HVPP (High Voltage Parallel Programming).

Another threat against MCUs and integrated circuits is the clock glitch attack. The processor and FPGA clock signals are often used to synchronize various parts of the circuit. In a clock glitch attack, the system clock frequency is intentionally increased for a short period of time. The glitched clock may disrupt CPU from correctly running instructions. When the clock is back to normal, later instructions will be executed correctly.
A clock glitch generator for fault injection attacks against an AES (Advanced Encryption Standard) circuit is designed and implemented on the SAKURA-G board in \cite{MSI::Clockglitch::GCCE::16}.
The SAKURA-G board is also used to implement a biased fault attack against a fault-resistant software implementation of LED block cipher to retrieve its secret key \cite{YGS+::Single-glitch::FDTC::16} . A detailed study is performed on the effects of clock glitches on legacy 8-bit AVR MCUs \cite{BGV::ClockGlitches::FDTC::11} . A glitchy-clock generator integrated in FPGA is presented for evaluating fault injection attacks and their countermeasures on cryptographic modules \cite{ESAS::glitchy-clock::JCE::11}. Clock-glitch attacks are also deployed under the impact of heating \cite{KHEV::ClockGlitches::FDTC::14}.

{\bf Operating system (OS)/firmware and software security and privacy}:
Given the often limited functionalities of an IoT device, a trustworthy operating system \cite{Parno:2011:BTM:2060078} can be implemented on the device if the cost is permitted. The control app for a thing is often installed on a smartphone and software secure measures should be applied in order to prevent the attack against the app like the attack in \cite{Kov::Belkin::2016}. We can also not blindly trust the cloud for security. For example, servers installed on Amazon EC2 have to be secured by whoever deploys the servers. Software security issues are similar to those in the traditional computer systems.
For example, backdoors and public and private SSL key pairs are discovered by performing static analysis on a large number of unpacked firmwares \cite{Costin+::StaticFirmwareAnalysis::Security2014}. Chen {\em et al.} \cite{Chen+::DynamicFirmwareAnalysis::NDSS2016} perform large-scale automated dynamic analysis of various firmwares to discover potential exploits using the Metasploit
Framework. A case study of a firmware modification attack is investigated in \cite{Cui+::FirmwareModificationsAttack::NDSS2013}. A buffer overflow exploit is found by analyzing Home Network Administration Protocol (HNAP)~\cite{TTYS0::plug::2014} so that it can be used to execute any code on the device. A stack-based buffer overflow of the general library, glibc~\cite{CVE::glibc::2015}, is exploited to attack several home hubs~\cite{Smith::SecurityHoles::2015}.

A MCU often does not have a full-fledged operating system, but a piece of dedicated code, i.e. {\em firmware}, for the particular application. The firmware has to be carefully designed and implemented to prevent malicious manipulation of the firmware. ATmega1284P's flash memory space is divided into two sections, application program section and boot loader section. A boot loader is a piece of code that runs when the chip is powered-up or restarted. For ATmega1284P, the boot loader can be designed to have three functions, {\em load firmware}, {\em boot firmware}, and {\em readback}. A physical jumper is used to put the boot loader into different functions. If an IoT device uses ATmega1284P, the load firmware and readback functions have to be secured in case that an attacker tries to hook up an ISP cable and read the firmware to compromise the intellectual property (IP). For example, readback requires a password.
ATmega1284P can store a crypto key in EEPROM. Therefore, a secure firmware distribution can be implemented. An encrypted firmware can be downloaded from the Internet. The boot loader can read the key in EEPROM, decrypt the encrypted firmware and write it into the flash.

{\bf Network Security and Privacy}:
An IoT system is a networked system and the whole system has to be secured from end to end \cite{Rouf+::Meter::CCS2012,Dhanjani::Lightbulbs::2013,Molina::KNX::DEFCON2014,Rahman::fitand::SP2013,Barcena::InsecurityIoT::2015,Obermaier::CameraSystems::IoTPTS2016,Zuo::BruteForce::NDSS2016}. Communication should be encrypted to prevent the leak of sensitive information. Authentication has to be carefully implemented. We have differentiated pairing from binding. Recall that in the pairing process, the controller needs to connect to the IoT device in order to configure the thing. However, most IoT devices allow any controller in proximity for pairing. The risk of such practice may be small in a private setting like a home. However, for a large-scale deployment in a public environment, anybody with access to the devices can reconfigure the system and may break into the system. After pairing, we run the binding process to bind identities to the thing in order to control it. The authentication has to be set up in a proper way. For example, weak passwords should be avoided. An IoT system may be composed of a large number of nodes with sensing capabilities and security techniques in sensor networks can be applied accordingly \cite{YHDZ::IoT::2013,XRS+::SensorKeyManagement::2007,DYGC::SensorKeyManagement::2007,DH::Sensor::2008}.

Many manufacturers fail to provide necessary protection for their networked IoT devices, which are under constant attacks nowadays. The Mirai DDoS attack \cite{Antonakakis+::MiraiBotnet::Security2017} was possible because of the weak passwords on various IoT devices. Rouf {\em et al.}~\cite{Rouf+::Meter::CCS2012} exploit the unsecured wireless communication protocol of automatic meter reading. Dhanjani~\cite{Dhanjani::Lightbulbs::2013} hacks the Phillips Hue lightbulb system and finds that the authentication mechanisms are not strong. Molina~\cite{Molina::KNX::DEFCON2014} exploits the KNX, a standardized home automation communication protocol, and finds that the lack of authentication and encryption allows an attacker to remotely control the appliances in a hotel. Rahman {\em et al.}~\cite{Rahman::fitand::SP2013} find the communication protocol vulnerabilities of the wearable device (Fitbit).
By automatically analyzing the applications and forging the authentication messages, Zuo {\em et al.}~\cite{Zuo::BruteForce::NDSS2016} design an authentication message generator to perform brute force attacks against the corresponding remote application server. Obermaier and Hutle \cite{Obermaier::CameraSystems::IoTPTS2016} investigate the vulnerabilities of communication protocols of four surveillance camera systems.

{\bf Big Data Analytics}:
Since the cloud sits between the controller and IoT devices, it can collect all the data. Many of the systems including Amazon AWS IoT are set up in this way. We have to question: should the cloud know everything and collect data about us and our belongings? For example, for remote authentication, should the cloud serve as the authentication server to authenticate controllers/things? However, big data collected by the cloud can help defeat attacks. For example, a proper intrusion detection system over the cloud can prevent another round of Mirai attack. Since things are often very specific, intrusion detection can be made easy.

\section{Protocols of Edmiax Cameras}
\label{sec::Protocol}

In this section, we present a case study of exploiting an IP camera system manufactured by Edimax under our view of IoT security and privacy. We first introduce the architecture of the camera system and then present the detailed communication protocol.

\subsection{Architecture of the Camera System}
\begin{figure}
  \centering
  \includegraphics[width=210pt]{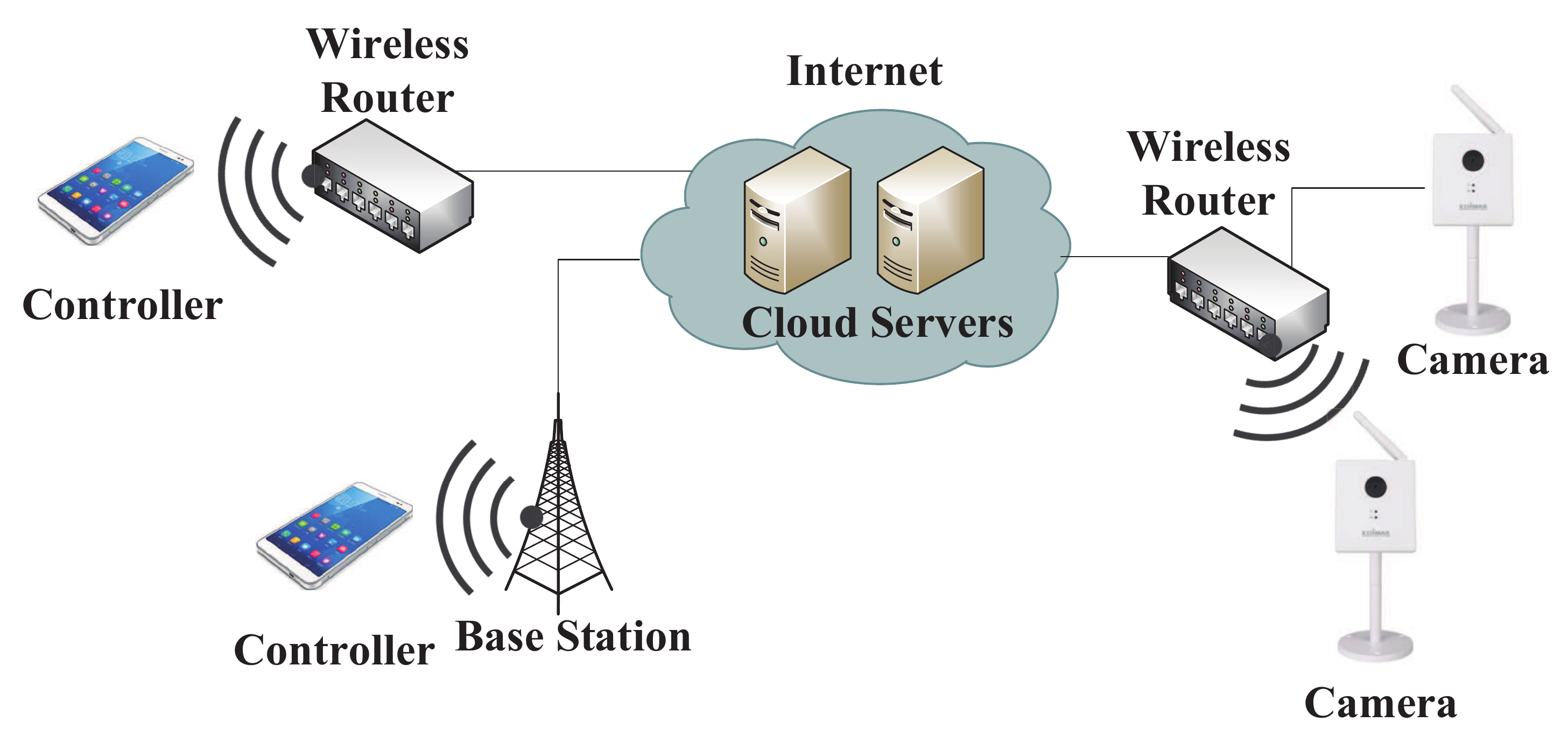}\\
  \vspace{-2mm}
  \caption{Architecture of the camera system}\label{fig::architecture}
  \vspace{-3mm}
\end{figure}

By traffic analysis, we find that the Edimax camera system has three components, including the camera, controller, and cloud servers as shown in Figure \ref{fig::architecture}.
If the controller and camera are in the same local network, the controller can communicate with the camera locally and fetch the live video through a web server on the camera. In this paper, we concentrate on remote attacks and will focus on remote communication protocols of the camera system when the controller and camera are not in the same local network.
The camera connects to the Internet through an ethernet cable or WiFi. The controller can be an app on a smartphone. The controller communicates with the camera via the cloud servers, including {\em the registration server} and {\em the command relay server}. The registration server is used for device registration for both the controller and the camera. The command relay server forwards command messages between them.

\subsection{Paring, Binding and Registration}

\begin{figure}
  \centering
  \includegraphics[width=200pt]{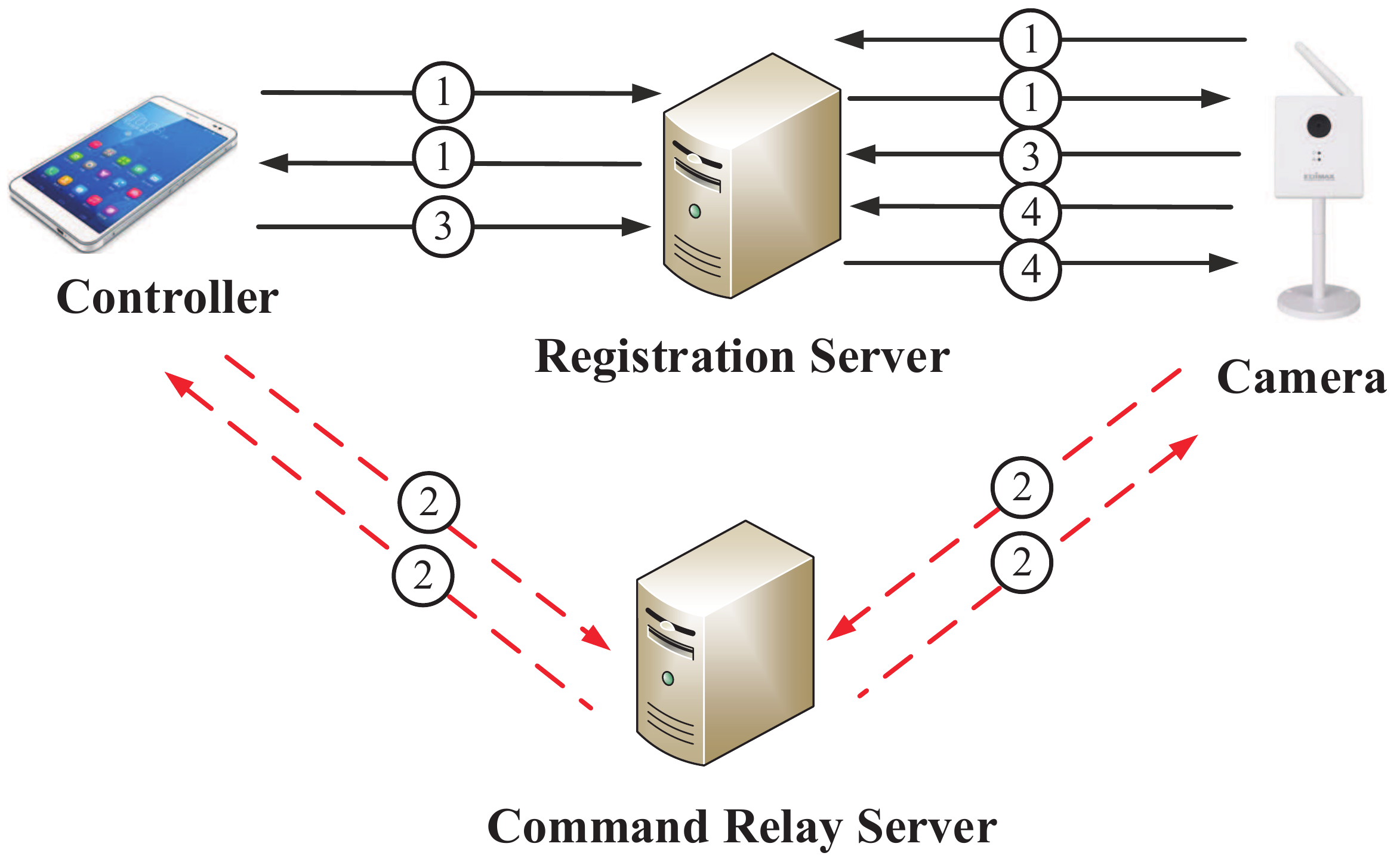}\\
  \vspace{-2mm}
  \caption{Registration phase}\label{fig::registration}
  \vspace{-3mm}
\end{figure}

We first investigate the paring process. When the camera is used for the first time, a user needs to connect it to her home network using an ethernet cable. The software {\it EdiView Finder Utility}  should be installed on a computer in the same home network. This utility is used to search the home wireless router and configure the camera to use the home wireless router. At this point, the wired connection of the camera can be disconnected.

In the binding process, we can change the password and other configurations such as the resolution of the image via a web page of the camera. The link is {\it http://host/setup.asp?r=20141126}, where {\it host} is the local IP address of the camera. Upon connecting to the Internet, the camera registers with two remote servers, i.e., registration server and command relay server. The controller also registers with both servers.

The packets transmitted between the controller, camera, and remote servers are obfuscated instead of encrypted. The right shift is performed over all characters but the first character in packets. The number of positions in the right shift is between 1 and 7, which is the difference between the first original character and the corresponding obfuscated character. The first original character is always ``$<$'', since a pair of ``$<>$'' is used to delimit key and value pairs. When the camera or controller receives a packet, it compares the first character with ``$<$'' to obtain the number of positions and perform the corresponding left shit to obtain the plaintext.

At the registration phase, all the packets use UDP. The UDP service ports of both the registration server and the command relay server are 8760. Figure \ref{fig::registration} illustrates the registration procedure for both the camera and controller. Since the first three steps of both the camera and controller are similar, we take the camera as an example to present the detailed procedure as follows.

\textbf{STEP 1:} In this step, the camera registers with the registration server. The camera first sends a UDP packet to the registration server. The packet has a value of ``1'' in the ``opcode value'' field, referred to as {\it command value} in this paper, and a UUID (Universally Unique Identifier) in the ``id value'' field. The UUID is used to uniquely identify the connection.

Upon receiving the packet with the command value ``1'' from the camera, the registration server responds with a UDP packet with a command value ``10''. The response packet consists of the UUID received from the camera, and the IP addresses and the ports of both the camera and the command relay server. Consequently, the camera can learn the IP address and port of command relay server from this response packet.

\textbf{STEP 2:} In this step, the camera registers with the command relay server. The camera first sends a UDP packet to the command relay server with a command value ``1'' and a new UUID to uniquely identify this connection. The command relay server responds with a UDP packet with a command value ``10''. The packet contains the UUID received from the camera, and the IP addresses and the ports of both the camera and the command relay server.

\textbf{STEP 3:} Once the camera receives response from the command relay server, it sends a packet with a command value ``2'' back to the registration server. This packet contains a new UUID. It is used to inform the registration server the fact that it has registered with the command relay server.

\textbf{STEP 4:} The camera sends two successive UDP packets to the UDP service port 8765 of the registration server. The first packet with a code value of ``3000'' is used to inform the registration server that the camera is online. The second packet with a code value of ``1010'' carries the camera information such as the camera model, MAC address, type, alias, LAN IP address and port of this camera, serial number, camera firmware version, and camera status.

After receiving the messages with the code value of ``1010'' from the camera, the registration server responds with a UDP packet with a code value of ``1020''. The packet contains the MAC address and the status of the camera. The camera repeats \textbf{STEP 1} to \textbf{STEP 4} around every 20 minutes to inform the registration server that the camera is online.

\subsection{Camera Discovery Phase and Authentication}

\begin{figure}
  \centering
  \includegraphics[width=200pt]{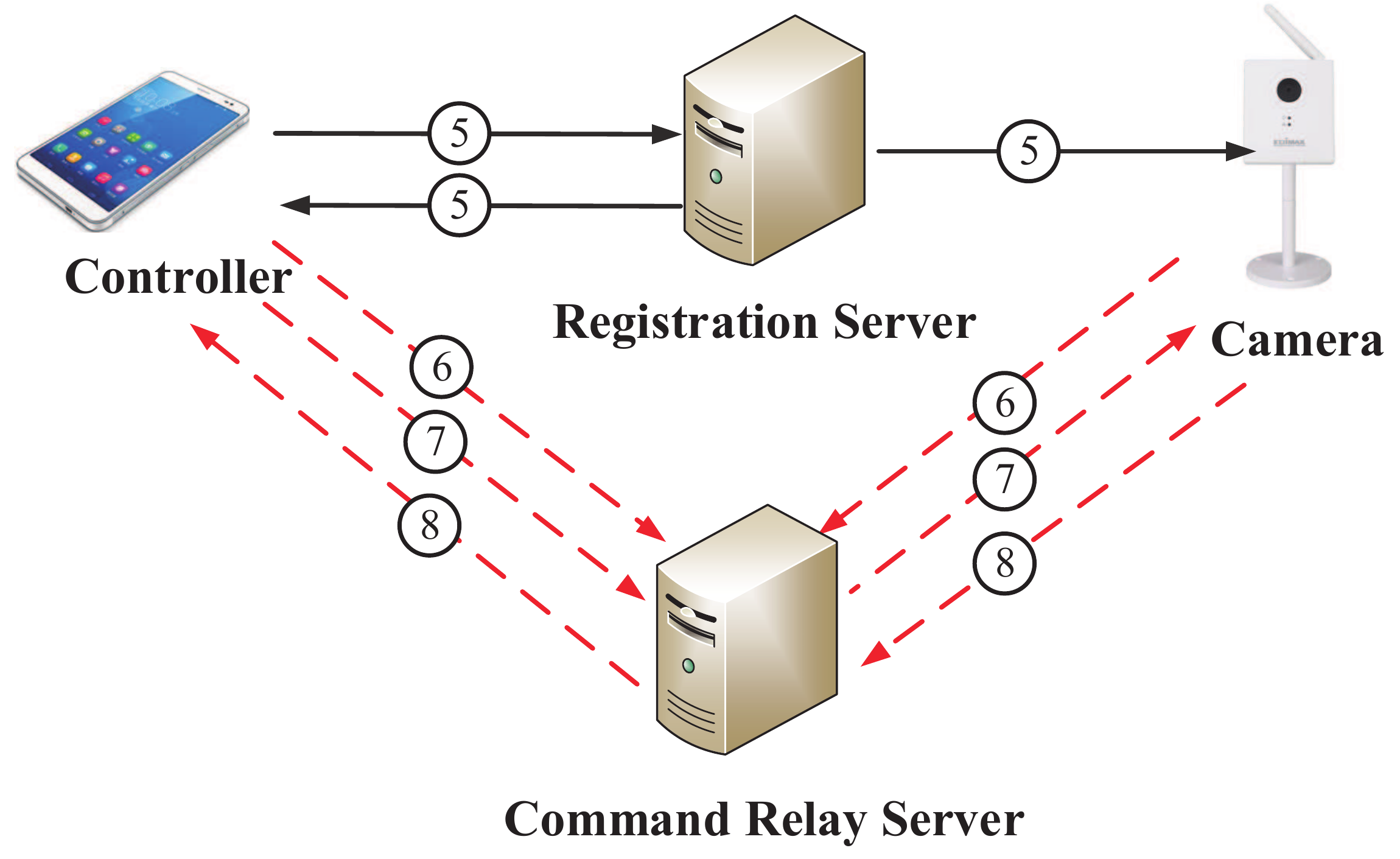}\\
  \vspace{-2mm}
  \caption{Connection establishment phase and data communication phase}\label{fig::connectionData}
  \vspace{-3mm}
\end{figure}

In the camera discovery phase, the controller tries to first check the online status of the camera via the registration server as shown in Figure \ref{fig::connectionData} and then sends the authentication information to it. The UDP service ports of the registration server for the camera and the controller are 8765 and 8766, respectively.

\textbf{STEP 5:} In this step, a user sets the configuration of the controller in order to check the online state of a specified camera. The user inputs an alias of the camera, the MAC address of the specified camera, and the password through the graphic user interface of the controller. The controller then sends two successive UDP packets to the registration server. The first packet with a code value of ``3000'' is to inform the registration server that the controller wants to check the state of the camera. The second packet with a code value of ``2030'' contains the MAC address of the camera and the information of the controller, including the LAN IP address and port, the device firmware version, and a relay ID generated by the controller. The relay ID is composed of the camera's MAC address and a timestamp. It is used in the data communication phase to correctly interconnect the two TCP connections from a pair of controller and camera on the command relay server.

After receiving the request from the controller, the registration server checks the camera status first. If the camera is offline, the registration server responds with a packet with a code value of ``5000''. Otherwise, the registration server responds with a packet with a code value of ``2040'' to the controller. This packet includes the IP addresses and ports of both the camera and the command relay server, the relay ID, camera firmware version, model, type, alias, and camera status. The registration server also adds extra messages to the ``2030'' packet and changes the code value to ``2020'', and then forwards it to the camera. The extra messages of ``2020'' packet include the IP addresses and ports of the camera, the controller, and the command relay server. Therefore, the camera can learn the relay ID from this packet.

\subsection{Remote Data Communication Phase}
There are two ways for the controller to control the camera remotely. First, the controller and the camera try to directly communicate with each other using the UDP protocol. 
Second, If the attempt of a direct UDP connection fails, the controller and the camera will communicate with each other via the command relay server using the TCP protocol. In this paper, we mainly concentrate on the data communication using TCP.

\textbf{STEP 6:} To communicate with TCP, both the camera and the controller establish TCP connections to the command relay server. Recall that the camera and the controller obtain the IP address and ports of the command relay server from the registration server. The camera also obtains the relay ID generated by the controller through the registration server. Both the camera and the controller send a TCP packet that contain the MAC address of the camera and the relay ID to the command relay server. According to the MAC address of camera and the relay ID, the command relay server can interconnect these two TCP connections and relay the data between the camera and the controller. However, the command relay server does not send any response packets to neither the camera nor the controller.

\textbf{STEP 7:} To obtain live images from the camera, the controller sends requests to the camera via the command relay server. The request packets contain a value of ``/mobile.jpg'' in ``url value'' field, and authentication information in ``auth value'' field. The authentication information in the format of {\it username}:{\it password} is encoded in the Base64 scheme. The default username and password are {\it admin} and {\it 1234}, respectively. Users can change the password through the web page of the camera. However, they cannot change the username as it is hardcoded in the camera. Once the command relay server receives the request packets from controller, it forwards them to the camera.

\textbf{STEP 8:} After the camera receives the request, the camera first checks the authentication information. If the authentication information is correct, the camera sends images back to the command relay server, which forwards them to the controller. Otherwise, the camera will send an authorization failure packet to the controller.

Every time the controller tries to obtain an image, it needs to send the request packet that contains the authentication information. Therefore, the controller repeats the \textbf{STEP 7} and \textbf{STEP 8} so as to continuously derive the live images taken by the camera.

\section{Security Vulnerabilities of Edimax Cameras}
\label{sec::vulnerabilities}

In this section, we first present three remote attacks against the Edimax IP camera of interest: device scanning attack, brute force attack, and device spoofing attack. Using these attacks, we can remotely control any camera. We then present various other attacks including local attacks. All the experiments were conducted over our own purchased cameras.

\subsection{Device Scanning Attack}
The attacker can find out all online cameras by enumerating all the possible MAC addresses. Recall the procedure of the connection establishment phase. After the controller sends a ``2030'' packet, the controller receives a ``2040'' packet if the camera is online. If the camera is offline, the controller will receive a packet with a code value of ``5000''. Therefore, the attacker can construct a ``2030'' packet with the specified camera MAC address, and check whether the specified camera is online according to the response packet.

The MAC address space of a manufacturer can be known from the Internet. A MAC address contains 12 characters. The first 6 characters indicate manufacturer and the other 6 characters indicate the namespace given to the manufacturer. Products of the same model from a manufacturer are usually assigned consecutive MAC addresses. Thus the attacker can infer MAC addresses based on the MAC address of his own purchased camera, enumerate the 12 characters of the MAC address and can verify the state of the camera with each potential MAC address.

\subsection{Brute Force Attack}
If a user changes the default password of a camera, the attacker can find the password via a brute force attack. In the data communication phase, when the controller sends a TCP request that contains the authentication information, the camera responds with images if the authentication information is correct. Therefore, the attacker can enumerate all possible passwords by repeating the TCP request, and determine if the password is right or not in terms of the response packet. Our experiments show that the command relay server does not block this brute force attack. If a user chooses a 4-digit password like the default one, the brute force attack works.

Although there is no explicit password policy from the manufacturer, we find that the camera password can be 63 characters long, and allows digits, special characters, upper-case, and lower-case alphabetic letters. Therefore, if the user employs a long and complicate password, the brute force attack may not work.

\subsection{Device Spoofing Attack}
The device spoofing attack can obtain a camera password of any length and combination. In the device spoofing attack, the attacker creates a software bot implementing the camera communication protocol in order to emulate the camera. When the user opens the control app, the TCP request packet with the password is sent to the attacker's software bot. Therefore, the attacker obtains the password.

The detailed attack process is presented as follows.
\begin{enumerate}
  \item The attacker chooses an online camera that uses a non-default password based on the device scanning results and creates the software bot with the specific MAC address. Any camera from this manufacturer can be spoofed this way.
  \item The software bot registers with the registration server and the command relay server by performing \textbf{STEP 1} to \textbf{STEP 4}. The software bot sends two UDP packets with the command value of ``1'' and ``2'' to both registration server and command relay server for registration. It then sends two successive UDP packets (i.e., code value of ``3000'' and code value of ``1010'') to the registration server informing the server that the spoofed camera is online. Once the software bot receives the packet with the code value of ``1020'' from the registration server, the attacker knows that the spoofed camera is online. The software bot repeats \textbf{STEP 1} to \textbf{STEP 4} as many times as possible, since the real camera will register itself by performing \textbf{STEP 1} to \textbf{STEP 4}.
  \item When the user opens the control app, the app sends two successive UDP packets (i.e., code value of ``3000'' and code value of ``2030'') to the registration server as introduced in \textbf{STEP 5}. The registration server forwards the packets to the software bot spoofing the camera. Simultaneously, the registration server informs the control app that the camera is online.
  \item The control app builds a TCP connection to the command relay server and sends a TCP request to the server automatically. The command relay server forwards the TCP request that contains the authentication information to the software bot. Recall that the authentication information is encoded with the Base64 scheme and the format is {\it username}:{\it password}. As a result, it is trivial for the attacker to derive the password from the authentication information.
  \item The spoofed camera should be offline as soon as it obtains the authentication information. Recall that the real camera registers with the registration server and the command relay server every 20 minutes. Accordingly, it takes at most 20 minutes for the real camera to get online again after the spoofed camera obtains the authentication information. After that, the user can see the images and videos taken by real camera again and may not realize that the camera has been compromised. Once the attacker obtains the password, she can fully control the camera.
\end{enumerate}

\subsection{Other Attacks}
\label{subsec::OtherAttacks}

We can perform various local attacks. 1. If the attacker can physically access the camera, she can click the reset button and perform the pairing using her own computer. Once pairing is done, she can bind the camera, e.g., changing the password of the camera. In this way, the attacker can fully control the camera. 2. If the controller and the camera are located in the same local network, the controller can directly access the live video via the web service provided by the camera. Their communicating traffic is in plaintext. Therefore,  a local attacker can trivially obtain the password when the user accesses the live video. 3. An attacker can install a malicious firmware into the camera via the link ``http://host/setup.asp?r=20141126''. In this way, the attacker can add various malicious software in the victim camera system and they can establish a reverse shell to a remote server so as to remotely execute commands in the camera as root.

We have also discovered a hidden backdoor vulnerability by using IDA Pro to manually reverse-engineer the CGI (Common Gateway Interface) scripts in the camera Linux operating system. There is a CGI script ``telnetd.cgi'' that can be used to start a telnet service. If an attacker is in the local network, she can start the telnet service by access the URL link {\em http://host/camera-cgi/private/telnetd.cgi?action=start}, where {\em host} is the local IP address of the camera and {\em action=start} is a pair of value and attribute passed to the CGI script. Authentication is required to start the telnet daemon while using the URL link. We can use the camera username and password obtained through various attacks, e.g., the device spoofing attack. Then the attacker can use the telnet username and password, ``admin" and ``1234", to log into the camera OS. Note that the username and password of the telnet service are not the ones used for accessing the live video from the camera, and cannot be changed by altering the password of the camera.

We can even remotely start the telnet service using the CGI script. Recall that a request used to obtain a live image from the camera contains a ``url value'' field in \textbf{STEP 7}. After we obtain the password of the camera via the device spoofing attack, we can send a request that contains a value of ``/camera-cgi/private/telnetd.cgi?action=start'' in the ``url value'' field to remotely start the telnet service. If a camera is deployed in a public network instead of behind a NAT router, we can remotely log into the camera system.

\section{Security Threats from Mirai}
\label{sec::propagation}
We have introduced the attacks against Edimax cameras in Section \ref{sec::vulnerabilities}. If those Edimax cameras use public IPs, we can install any malware including Mirai onto them. We have also showed that Edimax smart plugs suffer from device spoofing attacks \cite{LLX+::SmartPlug::IoTJ17}. Edimax smart plugs also suffer from a command injection attack in its password updating procedure and allow an adversary to run local commands such as {\em tftp}. Therefore an adversary can combine the device spoofing attack and command injection attack to download the Mirai malware and install it even if Edimax smart plugs are behind a wireless router. We have installed Mirai on our purchased smart plus this way. Given the IoT vulnerabilities discovered by us and other researchers, another wave of Mirai attack may affect us all in the foreseeable future.

In this section, we present our analysis of the released source code of the Mirai malware \cite{Antonakakis+::MiraiBotnet::Security2017} and explore the botnet architecture and propagation. The\nop{original and enhanced} propagation model is studied in detail. The goal is to understand the impact of Mirai.

\subsection{Mirai Botnet}
\begin{figure}
  \centering
  \includegraphics[width=250pt]{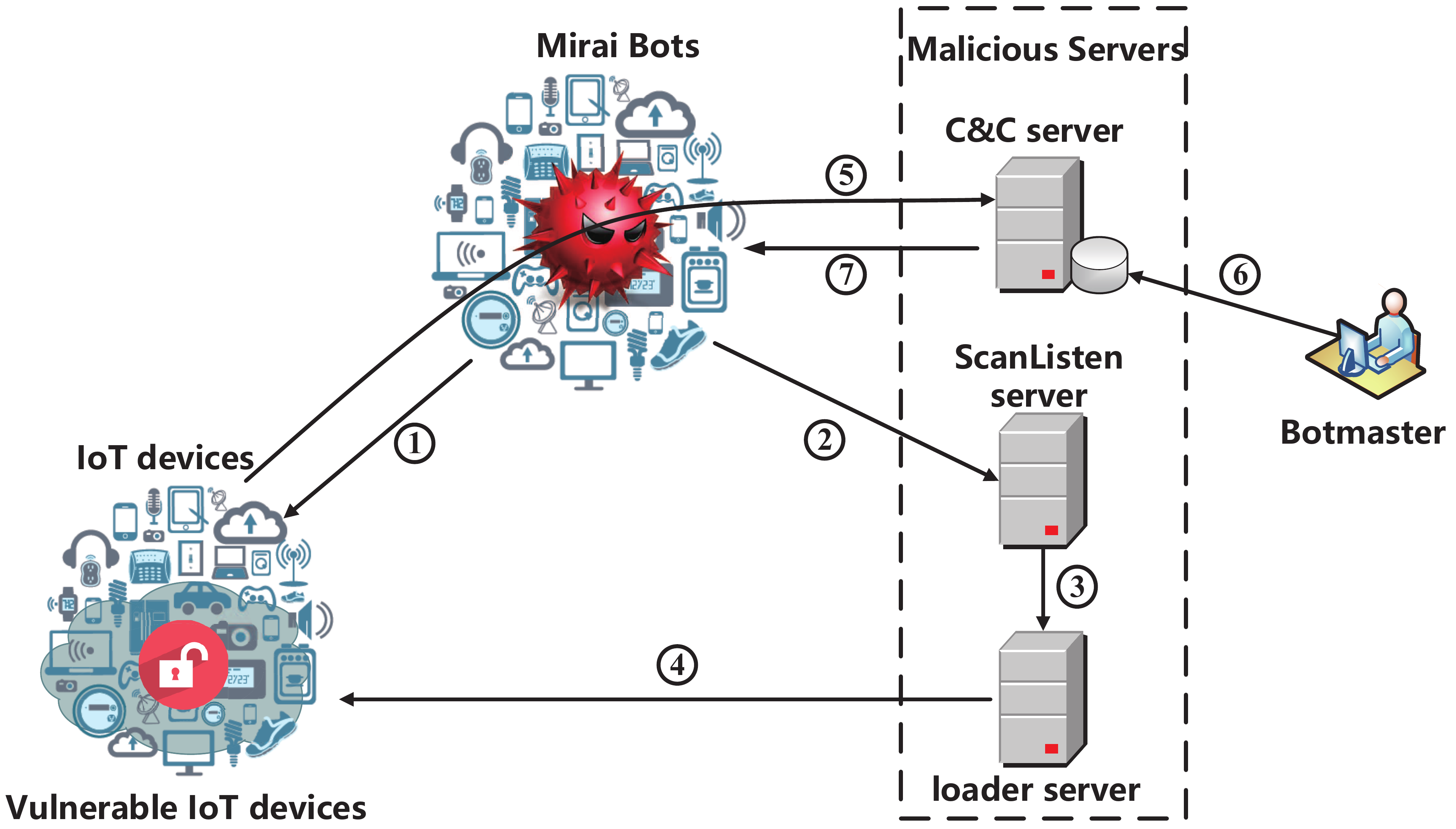}
  \caption{Propagation process of Mirai}\label{fig::MiraiCreatingBotnet}
\end{figure}

The Mirai botnet consists of four components as shown in Figure \ref{fig::MiraiCreatingBotnet}: bots, a C\&C (command and control) server, a ScanListen server, and loader servers. The bots are the IoT devices that are compromised by the Mirai malware and can continue to infect vulnerable devices by scanning the port of the Telnet service and deploying a password dictionary attack. The username/password list is hard-coded in the Mirai malware. The bots receive commands from a botmaster so as to conduct various attacks including distributed deny-of-service (DDoS) attacks. The C\&C server monitors the botnet status and the botmaster controls the bots via the C\&C server, e.g., sending attack commands. The ScanListen server receives information from newly discovered vulnerable IoT devices and forwards the information to the loader server. The loader server logs into a vulnerable device and infects it by executing the Mirai malware fetched from the loader server. The IoT devices can be divided into three categories: (1) IoT devices that close the Telnet service port, i.e., 23/2323 port. The bots never get a chance to establish TCP connections with these devices; (2) IoT devices that open the Telnet service port but their username/password is not in the username/password list of Mirai. The bots can perform the brute force attack but cannot compromise them; (3) IoT devices that open the Telnet service port and their username/password is in the list of Mirai. These devices can be compromised.

Figure \ref{fig::MiraiCreatingBotnet} illustrates the propagation procedure of Mirai. The detailed procedure is presented as follows.
\begin{itemize}
    \item {\bf Step 1 - Scanning}: The scanning starts with a single IP, which is treated as a bot for consistence. The bot scans and discovers vulnerable IoT devices over the Internet. The bot performs TCP SYN scans to probe the 23/2323 port of the public IP addresses. Mirai white-lists some IP addresses including those of the US Postal Service, the Department of Defense, the Internet Assigned Numbers Authority (IANA), and IP ranges belonging to Hewlett-Packard and General Electric. Upon obtaining a TCP SYN/ACK packet from an IoT device, the bot builds a TCP connection with the IP address of the device and then randomly chooses a pair of username/password from its dictionary  in an attempt to log into the Telnet server and executes commands to check whether the login is successful. By default, the dictionary has 62 pairs of username/passwords. However, the bot only tries at most 10 times on against a target.

    \item {\bf Step 2 - Bots reporting to  ScanListen server}:
    When a bot compromises a vulnerable IoT device, it reports the information of the device to the ScanListen server, including the IP address, port, and username/password of the target. The domain name and port of ScanListen server are also hard coded into the malware.

    \item {\bf Step 3 - Information relay from ScanListen server to loader server}: Once receiving the information of a compromised IoT device from the bot, the ScanListen server forwards the information of the device to the loader server.

    \item {\bf Step 4 - Loader server installing Mirai}: The loader server builds a TCP connection with the compromised device and logins with the received username/password. After successfully logging into the device, it first checks if a file transmission tool like {\em wget} or {\em tftp} is installed on the target system. If so, it downloads the Mirai malware from the loader server and execute the malware. Otherwise, the loader server connects to the telnet port of the target, reads a tiny binary file whose function is similar to the function of {\em wget}, and writes {\em echo} into the telnet connection. The string for {\em echo} is the hex characters from the binary file and the output stream of {\em echo} is redirected into a file, which will be saved at the target device.  Therefore, the binary file is transmitted from the loader server to the target as shown in Listing \ref{lst::echo}, and will be used to download the Mirai malware from the loader server.
        As soon as the Mirai is executed, it shuts down Telnet, SSH, and HTTP server on the compromised device to prevent other malwares or administrators from login. Mirai also binds on port 48101 to ensure only one instance of the malware is running on this device.

        {\scriptsize
        \begin{lstlisting}[caption={Loader server echoes a binary file into the target device},label={lst::echo}]
#define FN_DROPPER ``upnp".
#define TOKEN_QUERY ``/bin/busybox ECCHI".

 # Other code ...

util_sockprintf(conn->fd,
  ``echo -ne '%s' %s "FN_DROPPER "; ``TOKEN_QUERY``\r\n",
  conn->bin->hex_payloads[conn->echo_load_pos],
  (conn->echo_load_pos == 0) ? ``>" : ``>>");
        \end{lstlisting}
}
    \item {\bf Step 5 - Bot registration with C\&C server}: Once the Mirai malware is executed on a vulnerable device, the device establishes a TCP connection, registers with the C\&C server and becomes a member of the Mirai botnet. The domain name and port of the C\&C server are hard coded into the malware. Then the bot executes Step 1.

    \item {\bf Step 6 - Botmaster commanding}: A botmaster can send commands to the bots. The botmaster instructs the bots to launch an attack via C\&C server. Mirai is capable for multiple attacks, such as UDP DoS attack, TCP DoS attack, HTTP attack, and GREP attack.

    \item{\bf Step 7 - Botnet attacking}: The bots receive commands and start attack. Upon receiving the attack instruction from the C\&C server, the bots launch an attack on a target.
\end{itemize}


\subsection{Scanning Strategy}
We analyze released source code of the Mirai to study its scanning algorithm that is the key to the propagation speed. By analyzing the Mirai scanning algorithm in Algorithm \ref{alg::scanning}, we find that Mirai malware uses a uniform scanning strategy. Specifically, Mirai randomly scans public IP addresses and randomly selects a pair of username/password from a hardcoded list for the dictionary attack.
It uses the TCP SYN scanning technique to probe the 23/2323 port of a device to determine if the port is open. According to the Mirai source code, the bot first randomly selects 160 target IP addresses and uses a raw socket to send TCP SYN packets to these destinations.
Then it checks the raw socket 160 times in order to receive SYN+ACK packets.

Once a SYN+ACK packet is received, it builds a TCP connection with the IP address in the non-blocking mode and records the connection in a table. After checking and building connections using the raw socket 160 times, the bot checks the status of all TCP connections in the table. If a TCP connection is not established within 5 seconds, the bot gives up the connection request.

If the TCP connection to the telnet port of an IP is established successfully, the bot randomly tries a pair of username/password selected from a hardcoded list and also checks whether the device login is successful through the feedback from the target device. During this procedure, if the bot does not get any response from the device in 30 seconds, the bot will tear down this connection and establish a new TCP connection to guess a random username/password again. Mirai tries to crack the username/password of a target IP address up to 10 times. 

\begin{algorithm*}
\caption{Mirai Scanning Algorithm} \label{alg::scanning}
\begin{multicols}{2}
\begin{algorithmic}[1]
\REQUIRE  ~\\
(a) $\mu$ - the number of SYN probes sent by Mirai each time, \\
(b) $m$ - the size of a table, \\
(c) $table$ - TCP connection table, \\
(d) $table[i]$ - store the $ith$ TCP socket in $table$, \\ 
(e) $table[i].state$ - TCP connection state of the $ith$ entry of $table$, e.g., connecting or connected.
\ENSURE Discover the vulnerable devices
  \WHILE{TRUE}
    \IF {time interval since last SYN probe $>= 1s$}
		\STATE Send $\mu$ probes using a raw socket
	\ENDIF
	\WHILE{TRUE}
		\STATE Read SYN+ACKs packets from the raw socket
		\IF {error $||$ no data for reading $ || $ no entries in $table$}
			\STATE break
		\ELSIF{the $ith$ entry of $table$ is empty}
			   \STATE Build a connection, $table[i].state = connecting$
		\ENDIF
	\ENDWHILE
    \FOR { $i = 1:m $}
		\IF {the connection in $table[i]$ times out}
			\IF{ $table[i].state == connected$}
                \STATE re-connect(try times $<10$), otherwise free $table[i]$
			\ELSIF {$table[i].state == connecting$}
				\STATE free $table[i]$
            \ENDIF
		\ELSIF {$table[i].state == connecting$}
			\STATE Set the $table[i]$ in write file descriptor set (wr fdset)
		\ELSIF {$table[i].state == connected$}
			\STATE Set the $table[i]$ in read file descriptor set (rd fdset)
		\ENDIF
	\ENDFOR
    \STATE Select the write and read file descriptors
	\FOR {$i = 1:m $}
		\IF {$table[i]$ in wr fdset}
			\IF  {connection error for $table[i]$}
				\STATE free $table[i]$
			\ELSIF {connection success for $table[i]$}
				\STATE $table[i].state = connected$
			\ENDIF
		\ENDIF

		\IF {$table[i]$ in rd fdset}
			\WHILE {TRUE}
				\IF {connection error for $table[i]$}
                    \STATE re-connect(try times $<10$), otherwise free $table[i]$
					\STATE break
				\ELSIF { no data for reading}
					\STATE break
                \ELSIF {data for the last command}
                    \IF {login success}
						\STATE report to ScanListen server
					\ELSIF {login fail}
						\STATE re-connect(try times $<10$), otherwise free $table[i]$
                    \ENDIF
				\ELSE
                    \STATE send one command
				\ENDIF
			\ENDWHILE
		\ENDIF
	\ENDFOR
\ENDWHILE
	
\end{algorithmic}
\end{multicols}{2}
\end{algorithm*}

\subsection{Mirai Propagation Model}
According to the Mirai scanning strategy, we establish a Mirai propagation model to analyze the propagation speed. Unlike the malwares such as Code Red worm \cite{Zou+::CodeRedWormPropagation::CCS2002} that exploit software vulnerabilities, the Mirai malware not only scan the devices to discover the open 23/2323 port but also crack the username/password using a hardcoded dictionary so as to successfully infect IoT devices. We denote $q$ as the probability that a bot successfully finds the right pair of username/password. Since there are 62 pairs of username/password in the dictionary by default and the bot tries at most 10 times on a target, we can derive $q=[1-(61/62)^{10}]$.
Denote $\Omega$ as the total number of IP addresses, $N$ as the number of vulnerable IoT devices, and $\mu$ as the average scan rate of the bot. $I(t)$ is the number of bots at time $t$. At the start $t=0$, there are $I(0)$ bot(s) and $[N-I(0)]$ vulnerable devices. At time $t$, some vulnerable devices are discovered and infected, and we have $I(t)$ bots and $[N-I(t)]$ vulnerable devices. During a small time interval $\tau$, a bot can send $\tau \mu $ scans in total. The number of vulnerable IP addresses hit by the bot is $\frac{\tau\mu}{\Omega}[N-I(t)]$. Therefore, we can derive the number of compromised device during the time interval $\tau$ by
\begin{equation}
\label{eq::infectionRate}
\kappa =\frac{\tau\mu}{\Omega}[N-I(t)]q.
\end{equation}
The number of new bots created during the period between $t$ and $t+\tau$ is $\kappa I(t)$. Then we can obtain the number of bots at time $t+\tau$ by
\begin{equation}
I(t+\tau)=I(t)+\kappa I(t).
\end{equation}
Taking $\tau\rightarrow 0$ , we can have the Mirai propagation model by
\begin{equation}
\label{eq::propagation}
\frac{dI(t)}{d(t)}=\frac{\mu}{\Omega }[N-I(t)]I(t)q
\end{equation}

Note that $\kappa$ in Equation \eqref{eq::infectionRate} is affected by various factors, e.g., network congestion. $\kappa$ can decreases as the number of bots increases. 
Large-scare worm propagation may cause network congestion. 
It results in the reduction of the infection rate. Therefore, we change the number of compromised devices by a bot as follows:
\begin{equation}
\kappa (t)=\frac{\tau\mu}{\Omega}[N-I(t)][1-\beta \frac{I(t)}{N}]^{\alpha }q,
\end{equation}
where $\beta$ is the ratio of the number of vulnerable devices over the number of devices with 23/2323 port open. Recall that the bots will try the dictionary attacks against devices that open 23/2323 port while the telnet passwords are not in the dictionary. Such attacks generate traffic that contribute to the network congestion. $\alpha$ is used to adjust the infection rate sensitivity to the number of compromised devices. Then we can revise the propagation model of Equation \eqref{eq::propagation} by
\begin{equation}
\label{eq::propagationRevised}
\frac{dI(t)}{d(t)}= \frac{\mu }{\Omega }[N-I(t)][1-\beta \frac{I(t)}{N}]^{\alpha }I(t)q.
\end{equation}

\section{Evaluation}
\label{sec::Evaluation}

In this section, we present our experiment results validating the three attacks against Edimax IP cameras. All the attacks were performed over our own purchased cameras. In addition, we perform extensive simulations using NS3 \cite{NS3::NS3::2017} to evaluate the propagation speed of the original Mirai scanning strategy\nop{ and enhanced scanning strategy} in an ideal setup and also compare our theoretical modeling with real-world network telescope monitoring.

\subsection{Attack against Edimax IP Cameras}
To verify the feasibility of the device scanning attack, we first put our Edimax IP camera online. We then send a packet with a code value of ``2030'' to the Edimax registration server and receive a packet with a code value of ``2040''. We then put the camera offline. We resend a packet with a code value of ``2030'' to the registration server and receive a packet with a code value of ``5000''. Therefore, we can scan any potential MAC address to determine if the corresponding camera is online or not.

To verify the brute force attack, we set a random 4-digit password for our own camera. We then run the brute force attack and can identify the right password in a few minutes.

We now present the results of evaluating the device spoofing attack. The device spoofing attack may fail if the real camera registers with the registration server and this kicks our spoofed camera offline. In such a scenario, the controller will connect to the real camera, and the spoofed camera cannot receive the request packet with the authentication information from the controller. However, our software bot spoofing the camera can send out the registration packet continuously in order to increase the attack success rate. To verify our attack, we connect the real camera to the Internet, and the spoofed camera registers with the registration server every 10 seconds. A user opens the controller randomly during the attack. If the spoofed camera receives the authentication information, the attack succeeds; otherwise, it fails. We perform the experiments for 50 times and the spoofed camera receives the authentication information 49 times.The success rate of the device spoofing attack is up to 98\%.

\begin{figure}
  \centering
  \includegraphics[width=260pt]{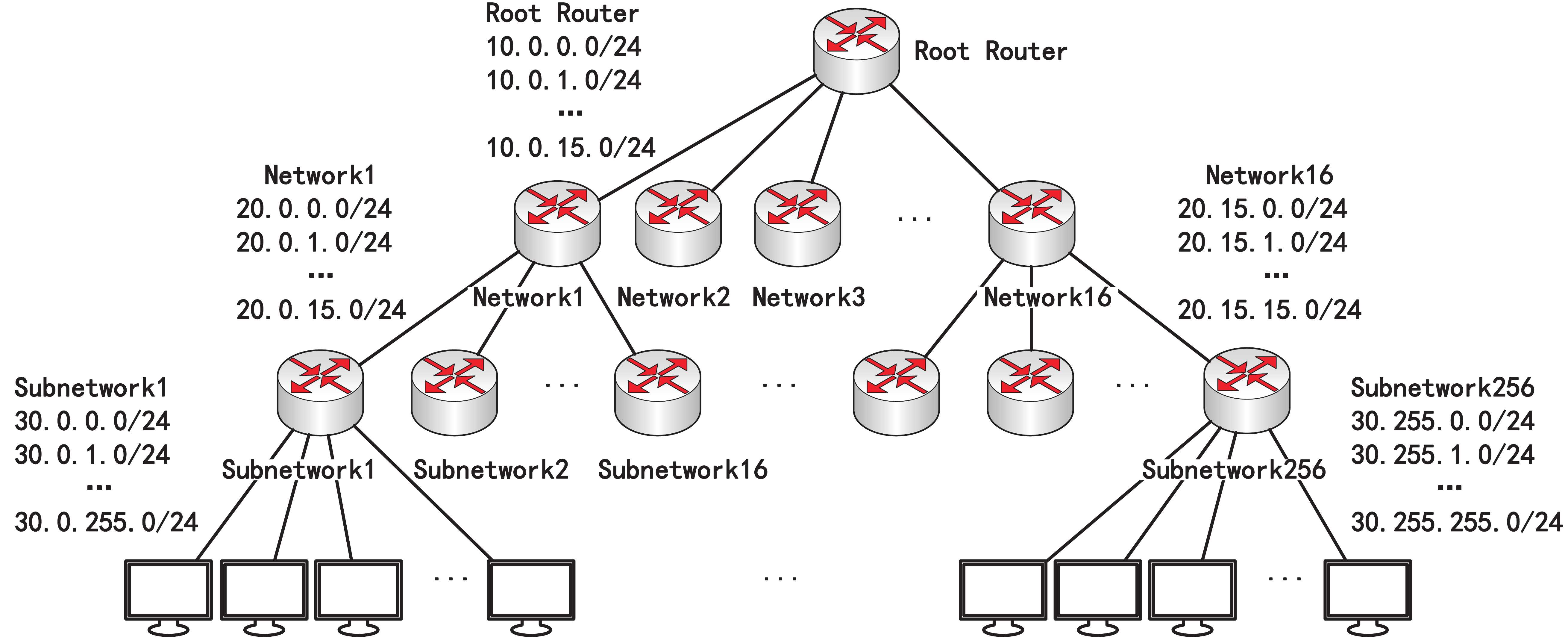}
  \caption{Network topology used in the simulation experiment}\label{fig::topology}
\end{figure}

\subsection{Mirai Propagation}
\label{subsec::Mirai}

We simulate the Mirai original propagation model using NS3.\nop{The goal of this set of simulations is to demonstrate that variants of Mirai could be seriously more vicious.} Recall that we divide the devices in the IP address space into three categories.
According to the results shown in ZoomEye \cite{ZoomEye::ZoomEye::2017}, 11.1\% online hosts have port 23 or 2323 open. Shodan \cite{Shodan::Shodan::2017} shows that 17.6\% of Telnet servers (23 port or 2323 port) use default credentials. Therefore, in our simulations, we assume 10\% of IoT devices in the IP space open port 23 and 20\% of the devices with the Telnet service are vulnerable. Assume the total number of devices in our simulated network is 65536. By using Monte Carlo method, we select 6325 (9.6\% of the total IP space) devices that open port 23 and 1270 (20.1\% of the Telnet servers) devices can be compromised by Mirai malware. Initially, there is one Mirai bot. Figure \ref{fig::topology} illustrates the network topology used in the simulation experiment. We use the NS3 global routing by building a global routing database for the topology using a Dijkstra Shortest Path First (SPF) algorithm. In addition, we set the throughput of devices and routers as 1000Mbps. The original Mirai scanning strategy is implemented in terms of Algorithm \ref{alg::scanning}. We repeat the simulation 100 times and derive the average number of bots $I(t)$ at time $t$.

\begin{figure}
  \centering
  \includegraphics[width=220pt]{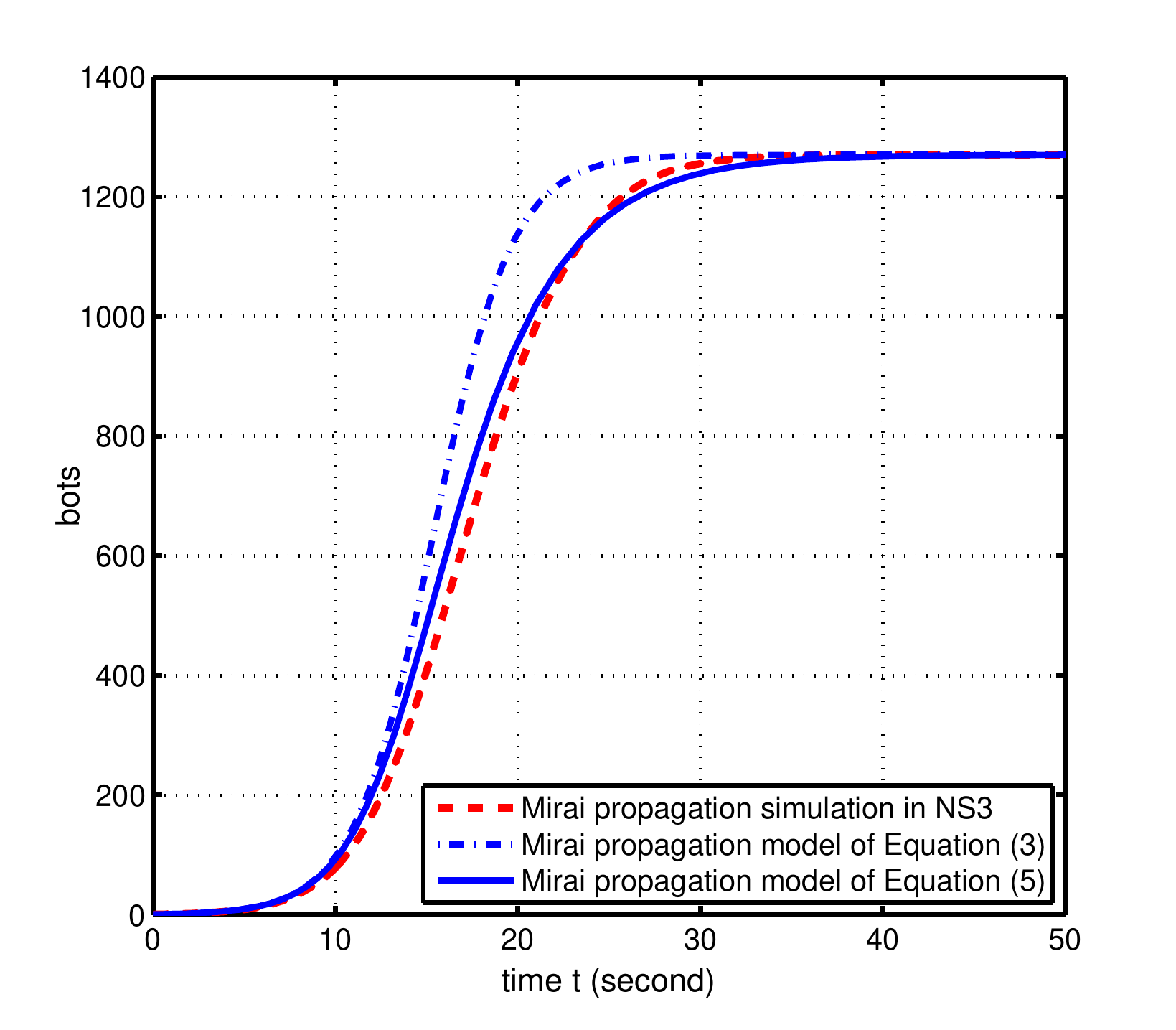}
  \caption{Simulation and theoretical results using the original Mirai propagation strategy}\label{fig::original}
\end{figure}

Figure \ref{fig::original} shows the relationship between $I(t)$ and $t$ in Equation \eqref{eq::propagation} and Equation \eqref{eq::propagationRevised} as well as the simulation results. 
It can be observed from Figure \ref{fig::original} that the theoretical curve obtained by Equation \eqref{eq::propagationRevised} matches the curve from the NS3 simulation result better. It demonstrates that $\kappa$ can be affected by network congestion as the number of compromised devices increases. NS3 simulation results show that 99\% of vulnerable devices can be compromised within 31 seconds. According to the theoretical curve obtained from Equation \eqref{eq::propagationRevised}, 99\% of vulnerable devices have be compromised within 34 seconds. So the theoretical result and simulation result are very close.

\begin{figure}
  \centering
  \includegraphics[width=220pt]{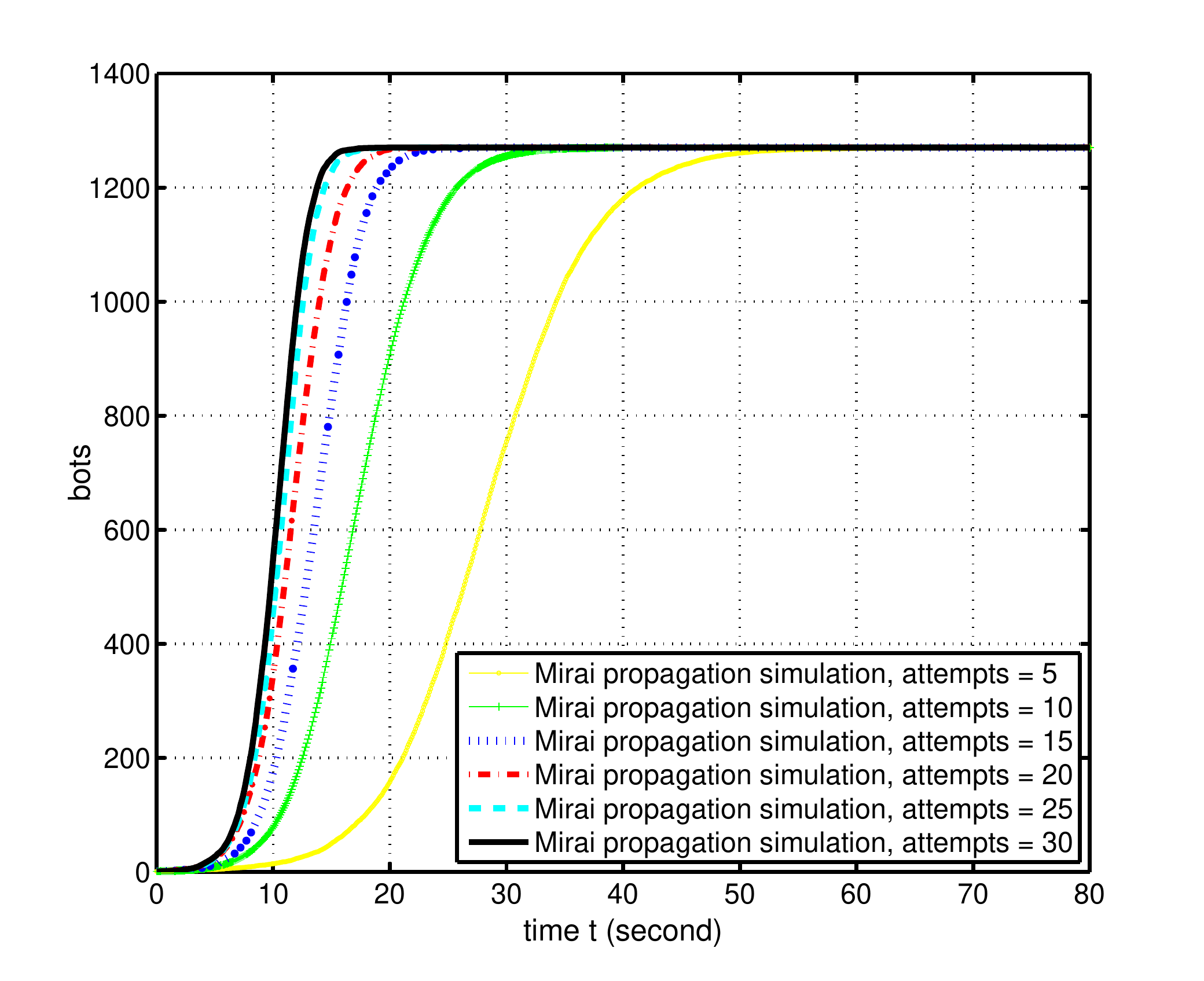}
  \caption{The Mirai propagation speed using different password cracking times}\label{fig::difftrytimes}
\end{figure}

To understand the impact of the number of username/password attempts on the Mirai propagation speed, we have performed simulations when the number of attempts is 5, 15, 20, 25 and 30 respectively. A pair of username/password is randomly chosen from its dictionary of 62 pairs of username/password every time. Once a correct username/password is selected, a bot will stop the attempt. We repeat the simulation 10 times and derive the average number of bots $I(t)$ at time $t$. Figure \ref{fig::difftrytimes} shows that as the number of attempts rises thus $q$ in Equation \eqref{eq::propagationRevised} increases, the propagation speed increases. Another observation is as the number of attempts increases, the gain (acceleration) of the propagation speed decreases. This can be caused by the extra generated traffic due to the increasing number of attempts (which will slow down the bot scanning rate) and congestion.

\nop{That is because the influence of $q$ in Equation \eqref{eq::propagationRevised} is smaller as the username/password try times increasing, however increasing the username/password try times also means increasing the username/password try times on IoT devices that open the Telnet service port but their username/password is not in the username/password list of Mirai, which will contribute to serious network congestion.}

\subsection{Comparison between Real-world Monitoring and Theoretical Modelling of Mirai Propagation}

\begin{figure}
  \centering
  \includegraphics[width=250pt]{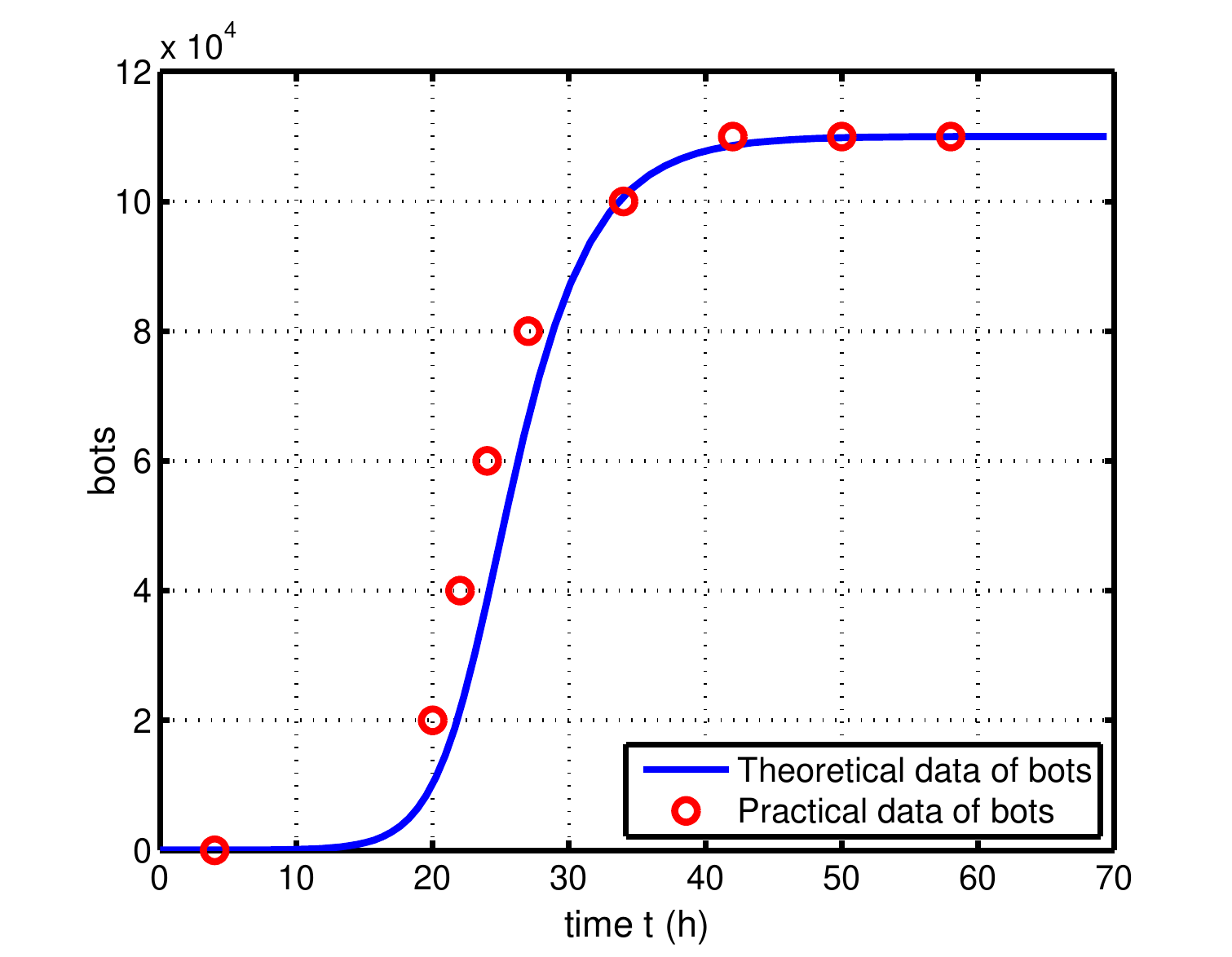}
  \caption{Comparison between theoretical results and practical data in \cite{Antonakakis+::MiraiBotnet::Security2017}}\label{fig::comparison}
\end{figure}

To demonstrate the effectiveness of our model, we also compare the results from our theoretical model of Equation \eqref{eq::propagationRevised} with the real-world network telescope results from Figure 4 in \cite{Antonakakis+::MiraiBotnet::Security2017}. Although there are data jitters in the practical curve in \cite{Antonakakis+::MiraiBotnet::Security2017}, the critical points in these two curves match well as shown in Figure \ref{fig::comparison}. As we can see from the results in \cite{Antonakakis+::MiraiBotnet::Security2017}, the number of vulnerable devices is around 110,000 during the first wave of Mirai attack. In our theoretical model in Equation \eqref{eq::propagationRevised}, we set $N$, $\Omega$, $\mu$, $\beta$, and $\alpha$ as 110,000, 3,417,112,576, 27, 0.2, and 3, respectively. Recall that $\alpha$  is used to adjust the infection rate that is affected by the number of compromised devices and thus congestion as a classical paper \cite{Zou+::CodeRedWormPropagation::CCS2002} on worm propagation shows. The scanning rate $\mu$ used in our model is 27 packets per second. Since the size of a TCP SYN scanning packet is 74 bytes, the scanning bandwidth is 1998 Bps. The scanning rate is reasonable according to the results from Figure 6 in \cite{Antonakakis+::MiraiBotnet::Security2017}. The long-time Mirai prorogation curve in \cite{Antonakakis+::MiraiBotnet::Security2017} is complicated because of the intrusion response effort from network administrators and various variants of Mirai introduced later.


%

\section{Conclusion}
\label{sec::Conclusion}

In this paper, we first present our view of an IoT system that includes the thing, cloud and controller from an end-to-end perspective. 10 basic functionalities have been identified for such a system. Those functionalities have to be secured properly according to our risk analysis of different components of the IoT system. We then present our exploit of an IP camera system and discovered three attacks including device scanning attack, brute force attack and device spoofing attack that can fully control all of the IP cameras from the manufacturer.
We performed real-world experiments to validate the attacks and find that the device spoofing attack can obtain a user's password at a probability of 98\% whatever the password is. Our end-to-end view of IoT Security and privacy can serve as the guide to design a secure and privacy preserving IoT system. The exploits of Edimax cameras in this paper and the exploits of Edimax smart plugs in our previous work allow the Mirai malware to be installed on those devices. To understand the potential impact of future Mirai attacks, we systematically model the propagation of Mirai\nop{and an enhanced version}. Our model and NS3 simulations match the trend of the Mirai propagation in real world.

Our work in this paper raises the alarm again for the IoT device manufacturers to better secure their products. The end-to-end view of IoT security and privacy can guide the effort of designing and protecting IoT systems.

\section*{Acknowledgments}

This work was supported in part by National Key R\&D Program of China 2017YFB1003000, National Natural Science Foundation of China under grants 61502100, 61532013, 61402104, 61572130, 61602111, 61632008, and 61320106007, by US NSF grants 1642124, 1461060, 1547428, 1723587, and Cisco, by Jiangsu Provincial Natural Science Foundation of China under grants BK20150637, by Jiangsu Provincial Key Technology R\&D Program under grants BE2014603, by Jiangsu Provincial Key Laboratory of Network and Information Security under grants BM2003201, by Key Laboratory of Computer Network and Information Integration of Ministry of Education of China under grants 93K-9 and by Collaborative Innovation Center of Novel Software Technology and Industrialization. Any opinions, findings, conclusions, and recommendations in this paper are those of the authors and do not necessarily reflect the views of the funding agencies.

\balance
\bibliographystyle{IEEEtran}
\bibliography{camera}

\end{document}